\documentclass[prl,superscriptaddress,twocolumn,showpacs]{revtex4}
\usepackage{graphicx}
\begin{document}
\def\i{{\rm i}}
\def\e{{\rm e}}
\title{The Leading Off-Diagonal Correction to the Form Factor of Large
  Graphs} 
\author{Gregory Berkolaiko}
\affiliation{The Weizmann Institute of Science, Rehovot 76100, Israel}
\author{Holger Schanz}
\email{holger@chaos.gwdg.de}
\affiliation{Max-Planck-Institut f\"ur Str\"omungsforschung und Institut
f{\"u}r Nichtlineare Dynamik der Universit{\"a}t G{\"o}ttingen, 
Bunsenstra{\ss}e 10, D-37073 G\"ottingen, Germany}
\author{Robert S. Whitney}
\affiliation{The Weizmann Institute of Science, Rehovot 76100, Israel}
\date{\today}
\begin{abstract}
  Using periodic-orbit theory beyond the diagonal approximation we
  investigate the form factor, $K(\tau)$, of a generic quantum graph
  with mixing classical dynamics and time-reversal symmetry.  We
  calculate the contribution from pairs of self-intersecting orbits
  that differ from each other only in the orientation of a single
  loop.  In the limit of large graphs, these pairs produce a
  contribution $-2\tau^2$ to the form factor which agrees with
  random-matrix theory. 
\end{abstract}
\maketitle
One of the fundamental open questions in the quantum theory of
classically chaotic systems is whether the fluctuations in the spectra
follow the predictions of random-matrix theory (RMT) \cite{BGS84} and
if so, precisely under which conditions. An interesting spectral
statistics to consider is the form factor $K(\tau)$, which is the
Fourier transform of the spectral two-point correlation function and
depends on a dimensionless time $\tau$ measured in units of the
Heisenberg time.  We will concentrate on systems with time-reversal
symmetry, the appropriate ensemble of random matrices is the Gaussian
orthogonal ensemble (GOE) and
the corresponding form factor is \cite{Haake}
\begin{eqnarray}
  \label{kgoe}
  K_{\rm GOE}(\tau)&=&2\tau-\tau\log(1+2\tau)\qquad(0\le \tau\le 1)
  \nonumber\\
  &=& 2\tau-2\tau^{2}+O(\tau^{3})\,.
\end{eqnarray}
There is ample numerical evidence \cite{Haake} for the conjecture that
in the semiclassical limit, Eq.~(\ref{kgoe}) correctly describes the
spectral fluctuations of any "typical" system with chaotic classical
limit, but there is no complete analytical theory. A natural starting
point for such a theory is the Gutzwiller trace formula \cite{Gut71}
expressing the spectral density in terms of classical periodic orbits
(POs). However, until recently, the only method for dealing with the
resulting double sum over POs was Berry's diagonal approximation
\cite{Ber85} which rests on the assumption that classical POs are
uncorrelated unless related by exact symmetries. This can be justified
only in the limit $\tau\to 0$, and the diagonal approximation results
in the leading-order term $2\tau$ but fails to reproduce any
higher-order corrections in Eq.~(\ref{kgoe}).

\begin{figure}[t]
  \def\fscale{0.44}
  \includegraphics[scale=\fscale]{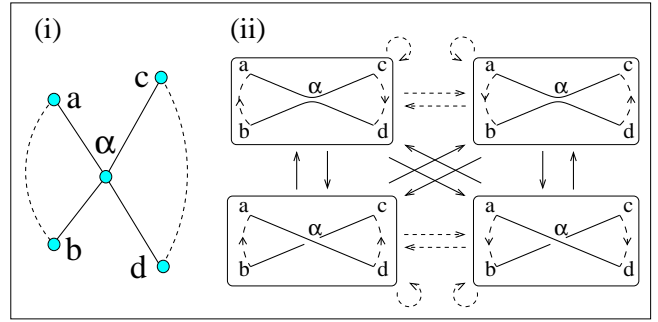} 
  \caption{In (i) we show two loops on
    a graph intersecting at a vertex $\alpha$.  The dashed arcs
    indicate trajectories between vertices $a$, $b$ and $c$, $d$,
    respectively. They may contain intermediate vertices which are not
    necessarily pairwise different.  In (ii) we show all orbits which
    traverse each of the loops exactly once.  Dashed arrows between
    orbits indicate pairs that were counted in the diagonal
    approximation.  Solid arrows indicate pairs missed by the diagonal
    approximation but counted by our method.}
  \label{fig:unfoldings8orbit}
\end{figure}

The analogy to diagrammatic perturbation theory for disordered systems
suggests that the leading correction to the diagonal
approximation is related to correlations between pairs of orbits which
differ only in one self-intersection \cite{AIS93,WLS99}
(see Fig.~\ref{fig:unfoldings8orbit}).  These correlations were recently
calculated for certain Riemannian surfaces with constant negative
curvature \cite{SR01,Sie01}, and they were found to reproduce the
$-2\tau^2$ term in Eq.~(\ref{kgoe}).
 
In this letter, we study the contribution from self-intersecting orbits in a
very different class of systems: quantum graphs \cite{KS97}.  In these systems
two-point correlation functions can be expressed as combinatorial sums and
therefore allow a treatment with fundamentally different analytical tools
\cite{SS99,BK99,Tan00}. Our result can be summarized as follows. The
contributions to the form factor from self-intersecting orbit pairs differing
in the orientation of a single loop are calculated analytically.  We find that
they add to $-2\tau^2$ if correlations in the {\em classical} dynamics of the
graph decay sufficiently fast (Eq.~(\ref{eq:mixing_condition_strong}) below).
The r\^ole of the unitarity of the quantum dynamics is particularly
transparent in our derivation and classical information enters only via the
Frobenius-Perron operator---a concept which is not specific to system or
representation. However, the most important aspect in which our result extends
Refs.~\cite{SR01,Sie01} is that it applies to {\em generic} graphs and not
only to uniformly hyperbolic systems where all POs share the same Lyapunov
exponent.

We consider a graph with $N$ vertices and $B$ directed bonds.
The quantum dynamics on the graph is defined by the energy dependent
unitary bond-scattering matrix $S$, which can be interpreted as the
discrete time-evolution operator on the $B$ dimensional Hilbert space
of directed bonds. The matrix element $S_{m'l,lm} =
\sigma^{(l)}_{m'm}\e^{\i kL_{ml}}$ prescribes a transition amplitude
from bond $m\to l$ to bond $l\to m'$.  This amplitude consists of a
phase $kL_{ml}$, where $L_{ml}$ is the length of the bond $m\to l$,
accumulated during the free propagation with wave number $k$ from
vertex $m$ to vertex $l$, and an element of the unitary
vertex-scattering matrix $\sigma^{(l)}$ describing the vertex $l$
\cite{KS97}. Matrix elements which do not correspond to any
bond-to-bond transitions are zero. Solely for the purpose of a
transparent presentation we assume that for each pair of vertices
$m,l$ there is at most one directed bond $m\to l$.  To ensure the
invariance of the system under time reversal we require the matrices
$\sigma^{(l)}$ to be symmetric and $L_{ml}=L_{lm}$ for all pairs of
connected vertices $l, m$. Otherwise different bond lengths are
assumed rationally independent.

The form factor of a graph can be defined as \cite{SS99}
\begin{equation}
  \label{ff}
  K^{(B)}(\tau)=B^{-1}\langle|{\rm tr}S^{t}(k)|^{2}\rangle_{k}
\end{equation}
where $\tau=t/B$ and the average is over the wave number
$\langle\cdot\rangle_{k}= \lim_{\kappa\to\infty}\kappa^{-1}
\int_{0}^{\kappa}{\rm d}k(\cdot)$.  We will be interested in the
limiting behavior $K(\tau) = \lim_{B\to\infty} K^{(B)}(\tau)$,
i.~e.~we first fix $\tau$ and consider larger and larger graphs, a
procedure corresponding to the semiclassical limit. Then we look at
the expansion of $K(\tau)$ around $\tau=0$.

Associated with the unitary matrix $S$ is the doubly stochastic matrix $M$
with $M_{m'l,lm} = |S_{m'l,lm}|^2 = |\sigma^{(l)}_{m'm}|^2$.  This matrix
defines a Markov chain on the graph which represents the classical analogue of
our quantum system \cite{KS97,BG00}. $M$ can be considered as the
Frobenius-Perron operator of this discrete classical dynamics. Physically,
$[M^t]_{m'l',ml}$ is the classical probability to get from bond $(l\to m)$ to
bond $(l'\to m')$ in $t$ steps.  We will assume the Markov chain to be mixing
(and consequently also ergodic), such that any initial distribution approaches
for long time an equidistribution over all bonds,
\begin{equation}\label{eq:mix}
\lim_{t\to\infty}M^{t}_{m'l',ml}={1\over B}
\end{equation}
for all pairs of bonds $m\to l$, $m'\to l'$. Eq.~(\ref{eq:mix}) is satisfied
if the eigenvalue unity of $M$, corresponding to the invariant
equidistribution, is the only eigenvalue on the unit circle while all others
are inside. For a given finite graph this is a very weak condition.  However,
we consider a family of graphs for which the limits of large graphs
$N\to\infty$ (and, as a consequence, $B\to\infty$) and large times
$t\to\infty$ are connected by $t/B=\tau$ being constant: the size of the graph
sets the time scale for the decay of classical correlations.  Therefore we
make Eq.~(\ref{eq:mix}) more precise and require as a {\em sufficient}
condition for the results to be derived below
\begin{equation}
  \label{eq:mixing_condition_strong}
  \lim_{B\to\infty}B^2\,N^3(B)\,\max
  \left|M^{\tau B}_{m'l',ml}-\frac1B\right|=0\,. 
\end{equation}
This is only slightly stronger than the recent conjecture by Tanner
\cite{Tan01}, that RMT behavior results in graphs if the spectral gap
between unity and the magnitude of the next biggest eigenvalue
vanishes slower than the inverse matrix size $1/B$. At the end of
this letter we give two examples, in which equidistribution is
approached at least exponentially fast as $B\to\infty$, such that our
results do apply.

Expanding the trace of the matrix powers in Eq.~(\ref{ff}) and
performing the average over $k$, we can represent the form factor in
terms of pairs of returning trajectories $p,q$ on the graph which
share the same length
$K^{(B)}(\tau)=B^{-1}\sum_{p,q}A_{p}A_{q}^{*}\delta_{L_{p},L_{q}}$
\cite{KS97,SS99}.  Here $p,q$ denote sequences of $t$ vertices
$p=[p_1,\cdots,p_t]$, such that for any $i$ there is a bond
$p_i\rightarrow p_{i+1}$.  The length of the corresponding trajectory
is $L_p=L_{p_1p_2}+L_{p_2p_3}+\ldots+L_{p_tp_1}$ and the amplitude is
$A_{p}=\sigma^{(p_1)}_{p_t,p_2}\sigma^{(p_2)}_{p_1,p_3}
\cdots\sigma^{(p_t)}_{p_{t-1},p_1}$.  If the trajectories $p,q$ share
the same length, then any pair of cyclic permutations of $p,q$ will do
so as well. All returning trajectories $p$ related by cyclic
permutations form an equivalence class $P$ which represents a periodic
orbit.  As $t\to\infty$, only an exponentially small fraction of the
set of orbits are multiple repetitions of shorter orbits, ignoring
these we write the form factor as a double sum over POs
\begin{equation}
  \label{posum}
  K^{(B)}(\tau)={t^2\over B}\sum_{P,Q}A_{P}A_{Q}^{*}\delta_{L_{P},L_{Q}}\,.
\end{equation}
Due to the rational independence of different bond lengths all contributions
to Eq.~(\ref{posum}) come from pairs of POs which make the same number of
transitions between any given pair of vertices $m,l$ (i.~e.\ $m\to l$ or $l\to
m$).  In particular, the pairs $P,P$ and $P,\overline P$, where bar denotes
the time-reversal, contribute to (\ref{posum}) for any $P$ since time-reversal
symmetry implies $L_{P}=L_{\overline P}$ (and $A_{P}=A_{\overline P}$).  These
contributions are collected in the diagonal approximation to the form factor
\begin{eqnarray}
  \label{diapp}
  K^{(B)}_{\rm diag}(\tau)
  &=& {t^2\over B} \sum_{P}(|A_P|^{2}+A_PA_{\overline P}^{*})\to 2\tau\,,
\end{eqnarray}
which reproduces the linear term in Eq.~(\ref{kgoe}).  To evaluate the sum
(\ref{diapp}) we used that according to Eq.~(\ref{eq:mix}) $\sum_{P}|A_P|^{2}
= t^{-1} {\rm tr} M^t \to t^{-1}$\,.  We also neglected self-retracing orbits
$P=\overline P$, since they represent an exponentially small fraction as
$t\to\infty$.

It is the purpose of this letter to go beyond the diagonal
approximation and investigate the contribution of pairs of orbits
which differ from each other at one self-intersection.  This is
illustrated by Fig.~\ref{fig:unfoldings8orbit}: (i) is the example of
two loops intersecting at a vertex $\alpha$.  (ii) shows the four
orbits which traverse each of the two loops once and thus have the
same length.  The arrows between these orbits indicate the possible
pairs whose contribution should be counted in (\ref{posum}); the
dashed arrows correspond to pairs already counted in the diagonal
approximation and the solid arrows to those that we have to sum here.
\begin{figure}[t]
  \def\fscale{0.45}
  \includegraphics[scale=\fscale]{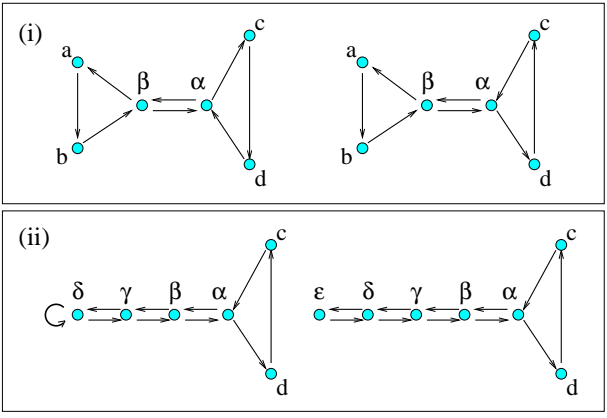} 
  \caption{Two conditions make the
    standard code unique: (i) If the vertex separating the two loops
    is ambiguous, we choose it such that the length of the loop with
    unchanged orientation is maximal, e.~g.\ in the top pair of orbits
    we take $\alpha$.  (ii) The loop with unchanged orientation must
    not be time-reversal invariant (self-retracing), e.~g.\ the bottom
    orbits are excluded.}
  \label{fig:sticky_orbit}
\end{figure}

The two orbits in any pair that is of interest to us differ in the orientation
along one of the cycles from vertex $\alpha$ to $\alpha$.  To count such pairs
we define a ``standard'' symbolic code for the orbits of the pair
$P=[p_1,p_2,\ldots,p_t]=[\alpha,l_1,\alpha,l_2]$ and
$Q=[\alpha,l_1,\alpha,\overline{l_2}]$, where $l_1=[p_{2},\dots,p_{t'-1}]$ and
$l_2=[p_{t'+1},\ldots,p_{t}]$ represent sequences of $t'-2$ and $t-t'$
vertices, respectively, and $\overline{l_2}=[p_{t},\dots,p_{t'+1}]$ stands for
the reversal of the sequence $l_2$.  Alternatively, a pair can also be
specified by the code of orbit $P$ and the position $t'$ of the second
occurrence of $\alpha$.  For example, the standard code for the pair of top
right and bottom right orbits of Fig.~\ref{fig:unfoldings8orbit}(ii) is
$P=[\alpha,a,\ldots,b,\alpha,d,\ldots,c]$ and
$Q=[\alpha,a,\ldots,b,\alpha,c,\ldots,d]$. We stress that the orbit $P$ will
in general contain more than one self-intersection. It is the pair $P,Q$ which
singles out the particular one at $\alpha$.  Nevertheless, the code for the
pair $P,Q$ is not yet unique. For example, the pair shown in
Fig.~\ref{fig:sticky_orbit} (i) has two possible codes:
\begin{eqnarray}
  P=[\alpha,\beta,a,b,\beta,\alpha,c,d],\ 
  Q=[\alpha,\beta,a,b,\beta,\alpha,d,c] & & \nonumber \\
\hbox{or }  P=[\beta,a,b,\beta,\alpha,c,d,\alpha],\ 
  Q=[\beta,a,b,\beta,\alpha,d,c,\alpha] & &\nonumber 
\end{eqnarray}
with $t'=6$ and $t'=4$, respectively (remember that cyclic permutations of the
code are irrelevant).  To avoid this ambiguity we require (i) $p_{t'+1}\neq
p_t$ for the standard code which excludes the second of the two
representations shown and requires $c\ne d$ for the first one to be valid.
Moreover we assume (ii) $l_1\ne\overline{l_1}$ since otherwise $Q=\overline P$
such that the orbit pair was already counted in the diagonal approximation.
Augmented with the conditions (i), (ii) the standard code is a unique
representation of all orbit pairs we have to count.  Thus we have the
contribution of such orbits to the form factor
\begin{eqnarray}
  \label{rsum_gen1}
  K_{\rm 1}^{(B)}(\tau)&=&
  {t^2\over B}\, \sum_{t'=4}^{t-2}{\sum_P}'
  (1-\delta_{cd})A_{P}A_{Q}^{*}
\end{eqnarray}
where the sum is over $P=[\alpha,a,p_3,\ldots,b,\alpha,c,\ldots,d]$
(in analogy to the notation of Fig.~\ref{fig:unfoldings8orbit} we write
$a,b,c,d$ for $p_2,p_{t'-1},p_{t'+1},p_t$). This sum is subject to the
restriction $l_{1}\ne \overline{l_1}$ indicated by a prime and also to
all restrictions implied by the connectivity of the graph (for some
sequence of vertices $P$ there might be no orbit).

Now we compute the contribution $A_PA_Q^*$ of an orbit pair $P,Q$.  The
amplitude $A_Q$ differs from $A_P$ only due to the vertex-scattering matrix
elements picked up at the self-intersection point $\alpha$. $P$ gets a
contribution $\sigma^{(\alpha)}_{da}\sigma^{(\alpha)}_{bc}$ from the two
encounters of vertex $\alpha$, while $Q$ has the combination of adjacent
vertices exchanged $\sigma^{(\alpha)}_{ca}\sigma^{(\alpha)}_{bd}$. Thus
\begin{equation}
  \label{eq:APAQ}
  A_{P}A_{Q}^{*}
  = |A_{l_1}|^2 |A_{l_2}|^2
  \sigma^{(\alpha)}_{da}\sigma^{(\alpha)}_{bc}
  \sigma^{(\alpha)*}_{ca}\sigma^{(\alpha)*}_{bd}\,,
\end{equation}
where $A_{l_i}$ is equal to the product of matrices $\sigma$ over the sequence
$l_i$. After inserting this result into Eq.~(\ref{rsum_gen1}) we can sum over
the variables $p_3,\ldots, p_{t'-2}$ and get $\sum_{p_3,\ldots,
  p_{t'-2}}|A_{l_{1}}|^2=M^{t'-2}_{\alpha a, b\alpha}$.  After a similar
summation over $p_{t'+2},\ldots, p_{t-1}$ we obtain
\begin{eqnarray}
  \label{rsum_gen2}
  K_{\rm 1}^{(B)}(\tau)&=&
  {t^2\over B}\, \sum_{t'=4}^{t-2}\ 
  {\sum_{\alpha,a,b,c,d}}M^{t'-2}_{\alpha a, b\alpha} 
  M^{t-t'}_{\alpha c, d\alpha} (1-\delta_{cd})\nonumber\\
  && \times
  \sigma^{(\alpha)}_{da}\sigma^{(\alpha)}_{bc}
  \sigma^{(\alpha)*}_{ca}\sigma^{(\alpha)*}_{bd}
  -K^{(B)}_{\rm 1, srt}(\tau)\,.
\end{eqnarray}
Performing an unrestricted summation over $p_3,\ldots, p_{t'-2}$ we
lost track of the condition $l_{1}\ne \overline{l_1}$, therefore the
contribution from such self-retracing orbits is removed by
subtracting $K^{(B)}_{\rm 1, srt}(\tau)$ which will be calculated
below.

Note that for large $t$ in Eq.~(\ref{rsum_gen2}) either $t'-2$ or $t-t'$ will
be large for any $t'$.  Thus we partition the sum over $t'$ into two
intervals, $\sum_{t'=4}^{t-2}=\sum_{t'=4}^{[t/2]}+ \sum_{t'=[t/2]+1}^{t-2}$,
where $[\cdot]$ stands for the integer part. We use Eq.~(\ref{eq:mix}) to
argue that in the first interval $M^{t-t'}= 1/B$, while in the second interval
$M^{t'-2}= 1/B$.  Calling $K_{\rm 1,S(L)}$ the contribution of the interval
with Small (Large) $t'$, we have $K_1= K_{\rm 1,S}+K_{\rm 1,L}-K_{\rm 1,
  srt}$.  Substituting $M^{t'-2}_{\alpha a, b\alpha}=1/B$ into $K_{\rm 1,L}$
we evaluate the sum over $a$ and use the unitary of $\sigma^{(\alpha)}$ to
show that $K_{1,L}=0$.  There are corrections to this result due to the
deviation from equidistribution of $M^{t'-2}$.  However so long as the
stronger mixing condition given by Eq.~(\ref{eq:mixing_condition_strong})
holds these corrections can be shown to vanish as $N\to\infty$.  To show this
we use the unitarity of $\sigma$ and the fact that all matrix elements of
$M^{t-t'}$ are bounded from above by unity.

For $K_{\rm 1, S}$  we set $M^{t-t'}_{\alpha c, d\alpha}=1/B$, then
the sum over $c$  can be written as
$
  \sum_{c} \sigma^{(\alpha)}_{bc} 
  \sigma^{(\alpha)*}_{ca} (1-\delta_{cd})= 
  \delta_{ab} - \sigma^{(\alpha)}_{bd}\sigma^{(\alpha)*}_{da}
$, 
since the matrix $\sigma^{(\alpha)}$ is both unitary and symmetric.
Hence $K_{\rm 1, S}$ simplifies to
\begin{eqnarray}
 \label{eq:K-short}
  K_{\rm 1, S}^{(B)}
  &=&{t^2\over B^2}\sum_{t'=4}^{[t/2]}\ 
  \left( \sum_{\alpha,a,d} M^{t'-2}_{\alpha a, a\alpha}
    |\sigma^{(\alpha)}_{da}|^2\right.\nonumber\\
  &&\qquad \left. 
  - \sum_{\alpha,a,b,d}
  M^{t'-2}_{\alpha a, b\alpha} |\sigma^{(\alpha)}_{da}|^2
  |\sigma^{(\alpha)}_{bd}|^2\right).
\end{eqnarray}
In the first sum inside the brackets we can sum $|\sigma^{(\alpha)}_{da}|^2$
over $d$ to produce one.  In the second sum we substitute
$|\sigma^{(\alpha)}_{da}|^2 = M_{d\alpha, \alpha a}$ and
$|\sigma^{(\alpha)}_{bd}|^2 = M_{b\alpha, \alpha d}$ and then use $\sum_{a,b}
M_{d\alpha, \alpha a} M^{t'-2}_{\alpha a, b\alpha} M_{b\alpha, \alpha d} =
M^{t'}_{d\alpha,\alpha d}\,.  $ We arrive at
\begin{eqnarray}
  &&K_{\rm 1, S}^{(B)}={t^2\over B^2}\sum_{t'=4}^{[t/2]}
  \sum_{a,b} \left(M^{t'-2}_{ab, ba} 
    - M^{t'}_{ab, ba}\right)\nonumber\\
  \label{eq:res_uncorrected}
  &&= {t^2\over B^2} \sum_{a,b} \left( M^2 + M^3 
    - M^{[t/2]-1} - M^{[t/2]}\right)_{ab, ba}\!\!.
\end{eqnarray}
Finally we compute the contribution of $K^{(B)}_{\rm 1, srt}(\tau)$ to
Eq.~(\ref{rsum_gen2}) from orbits with $l_{1}=\overline{l_1}$
(Fig.~\ref{fig:sticky_orbit}(ii)).  Such orbits can be described as follows:
To the left of vertex $\gamma$ they must have the same structure as one of the
orbits in Fig.~\ref{fig:sticky_orbit}(ii).  To the right of $\gamma$ they can
take any path which comes back to $\gamma$ (with the exception of {\em
  completely\/} self-retracing paths which are exponentially few in number and
can therefore be ignored). The sum over all these configurations to the right
of $\gamma$ yields $M^{t-2}_{\delta\gamma,\gamma\delta}$ or
$M^{t-3}_{\delta\gamma,\gamma\delta}$ for even or odd $t'$, respectively.  The
contribution of the transitions from $\gamma$ to the left can also be
expressed in classical terms, resulting in
\begin{equation}
  \label{srt}
  K_{\rm 1, srt}^{(B)}=
  {t^2\over B}\, \sum_{\gamma,\delta}\left(
    M^{2}_{\gamma\delta,\delta\gamma}M^{t-2}_{\delta\gamma,\gamma\delta}
    +M^{3}_{\gamma\delta,\delta\gamma}M^{t-3}_{\delta\gamma,\gamma\delta}
  \right).
\end{equation}
Matrix elements of large powers of $M$ can again be approximated by
$B^{-1}$ both above and in Eq.~(\ref{eq:res_uncorrected}).  Then we
see that the first two terms in Eq.~(\ref{eq:res_uncorrected}) are
exactly canceled by $K_{\rm 1, srt}^{(B)}(\tau)$.  The remaining terms
in Eq.~(\ref{eq:res_uncorrected}) give our main result
\begin{eqnarray}
  \label{eq:result}
  K_1(\tau)&=&-2\tau^2\,.
\end{eqnarray}
Eqs.~(\ref{diapp}), (\ref{eq:result}) show that the contributions to the form
factor of the orbits discussed here precisely reproduce the first two terms of
the RMT result, Eq.~(\ref{kgoe}).  Note, however, that we have not discussed
whether other correlated orbits could produce contributions of the same or
lower order.  This, we believe, can be shown by considering more and more
complicated pairs of orbits, eventually deriving a complete expansion of the
form factor in powers of $\tau$.

Finally we briefly discuss two specific models, to which
Eq.~(\ref{eq:result}) applies. We consider complete graphs with $N$
vertices and $B=N^2$ directed bonds connecting every pair of vertices
including a bond from a vertex to itself.  The first model we consider
has ``discrete Fourier transform'' (DFT) boundary conditions at each
vertex \cite{Tan01}. The second has Neumann boundary conditions \cite{KS97}
\begin{equation}
  \sigma_{mn}^{(l)}=
  \left\{\begin{array}{ll}N^{-1/2}\e^{2\pi\i mn/N}/\sqrt{N}&\mbox{(DFT)}\\
      2N^{-1}-\delta_{mn}&\mbox{(Neumann)}
    \end{array}\right.\,.
\end{equation}
The DFT graph is a uniformly hyperbolic model: the weights of all classical
orbits with the same length are equal. We have $M^{t}_{ab,cd}\equiv 1/N^2$ for
$t>1$, i.~e.\ all correlations vanish immediately and Eq.~(\ref{eq:mix}) is an
identity. For the Neumann graph, which is not uniformly hyperbolic, the
situation is more complicated: Here we have
$M^{2T}_{ab,cd}=F_{T}(a,c)F_{T}(b,d)/N^{2}$ and
$M^{2T+1}_{ab,cd}=F_{T+1}(a,d)F_{T}(b,c)/N^{2}$ with
$F_{t}(a,b)=1+(\delta_{ab}N-1)(1-4/N)^t$, but substituting $t=N^{2}\tau$ and
using $(1-4/N)^{N}\to\e^{-4}$ as $N\to\infty$ it is easy to see that $M^{t}$
converges for large graphs exponentially fast to equidistribution.

Inspiring discussions with M.~Sieber and U.~Smilansky are gratefully
acknowledged. HS thanks the WIS Rehovot for kind hospitality.  GB was
supported by the Israel Science Foundation, a Minerva grant, and the Minerva
Center for Nonlinear Physics. RW was supported by the U.S.-Israel Binational
Science Foundation (BSF) and the German-Israel Foundation (GIF).


\newpage
\appendix
\section{Appendix: Derivation of Eq.~(\ref{eq:mixing_condition_strong})}

We include this appendix (which is only in the nlin.CD version 
of this paper) in the hope that it will interest the reader.
In the paragraph below Eq.~(\ref{rsum_gen2}) we noted that if the condition
in Eq.~(\ref{eq:mixing_condition_strong}) is satisfied then $K_{1,L}=0$.
Here we show how we arrive at this condition.

We can write $K_{1,L}$ as
\begin{eqnarray*}
  \label{eq:correction}
  K_{1,L} = {t^2\over B}\sum_{t'=[t/2]+1}^{t-2} &&\hspace{-5mm} 
  \sum_{\alpha,a,b,c,d}
  M^{t-t'}_{\alpha c, d\alpha}
  \left(M^{t'-2}_{\alpha a, b\alpha} - \frac1B\right) \nonumber\\
  &\times&
  \sigma^{(\alpha)}_{bc} \sigma^{(\alpha)*}_{bd}
  \sigma^{(\alpha)}_{da} \sigma^{(\alpha)*}_{ca}  
  (1-\delta_{cd}).
\end{eqnarray*}
To estimate it from above we notice that 
$\max\left|M^t_{\alpha c, d\alpha}\right|$ decreases with $t$ (we take
the maximum over $\alpha,c,d$); and so does 
$\max\left|M^t_{\alpha a, b\alpha} - 1/B\right|$.  Thus
\begin{eqnarray}
  \label{eq:estimation_of_correction1}
  |K_{1,L}| \leq \frac12 B^2\tau^3 \max\left|M^2_{\alpha c, d\alpha}\right|
  \max\left|M^{[t/2]-1}_{\alpha a, b\alpha} - \frac1B\right|
  \nonumber\\
  \times\sum_{\alpha,a,b,c,d} 
  \left|\sigma^{(\alpha)}_{bc} \sigma^{(\alpha)*}_{bd}
    \sigma^{(\alpha)}_{da} \sigma^{(\alpha)*}_{ca}\right|.
\end{eqnarray}
Next we apply Cauchy-Schwartz inequality to the sum over $a$
\begin{equation}
  \label{eq:schwartz}
  \sum_{a} \left|\sigma^{(\alpha)}_{da} 
    \sigma^{(\alpha)*}_{ca}\right|
  \leq
  \left(\sum_a \left|\sigma^{(\alpha)}_{da} \right|^2 \right)^{\frac12}
\left(\sum_a \left|\sigma^{(\alpha)}_{ca}\right|^2 \right)^{\frac12}
  = 1
\end{equation}
Applying the same inequality to the sum over $b$ we obtain
\begin{equation}
  \label{eq:estimation_of_correction2}
  |K_{1,L}| \leq \frac12 B^2\tau^3 \max\left|M^2_{\alpha c, d\alpha}\right|
  \max\left|M^{[t/2]-1}_{\alpha a, b\alpha} - \frac1B\right|
  \sum_{\alpha,c,d} 1,
\end{equation}
where the last sum is over all vertices $\alpha$ and over vertices
$c,d$ adjacent to $\alpha$.  Thus to ensure that $|K_{1,L}|\to 0$ as we take
the limit $B,N,t\to\infty$ ($\tau = t/B$ fixed) we demand that
\begin{equation}
  \label{eq:almost_result}
  B^2 \max\left|M^2_{\alpha c,
  d\alpha}\right|\max\left|M^{B\tau/2}_{\alpha a, b\alpha} -
  \frac1B\right| \sum_\alpha v_\alpha^2 \to 0,
\end{equation}
where $v_\alpha$ is the valency of vertex $\alpha$. 
To make the estimate less dependent on the graph's topology we note
that $\left|M^2_{\alpha c,d\alpha}\right|<1$ and $v_\alpha \le N$,
then
\begin{equation}
  \label{apdx-eq:result}
  B^2 N^3 \max\left|M^{\tau B}_{\alpha a, b\alpha} - 1/B\right| \to 0,
\end{equation}
as $B$ (and $N$) tend to infinity.
Thus we arrive at the condition we gave in 
Eq.~(\ref{eq:mixing_condition_strong}).
Note a similar calculation shows that this is also a {\it sufficient} 
condition for corrections to Eq.~(\ref{eq:K-short}) to vanishes.

\end{document}